\documentclass[preprint,3p,twocolumn,authoryear]{elsarticle}

\usepackage[version=3]{mhchem} 
\usepackage{units}
\usepackage{amssymb}

\journal{}

\usepackage{etoolbox}

\makeatletter 
\def\ps@pprintTitle{  
\let\@oddhead\@empty 
\let\@evenhead\@empty  
\def\@oddfoot{\hfill\thepage}  
\def\@evenfoot{\thepage\hfill}} 
\makeatother

\usepackage{color}
\usepackage{hyperref, soul}
\definecolor{linkColor}{rgb}{1,0,0}
\definecolor{citeColor}{rgb}{1,0,0}
\hypersetup{pdfborder={0 0 0}, colorlinks=true,urlcolor=linkColor,citecolor=citeColor}

\begin{document}

\begin{frontmatter}

\title{Fabrication of Specimens for Atom Probe Tomography using a Combined Gallium and Neon Focused Ion Beam Milling Approach }

\author[label1,label2]{Frances I. Allen}
\ead{francesallen@berkeley.edu}

\address[label1]
{Department of Materials Science and Engineering, UC Berkeley, Berkeley, CA 94720, USA}
\address[label2]
{National Center for Electron Microscopy, Molecular Foundry, LBNL, Berkeley, CA 94720, USA}

\author[label3]{Paul T. Blanchard}
\author[label3]{Russell Lake}
\author[label3]{\texorpdfstring{David Pappas\fnref{fn1}}}

\address[label3]
{Physical Measurement Laboratory, NIST, Boulder, CO 80305, USA}

\fntext[fn1]{Current address: Rigetti Computing Inc., Berkeley, CA 94710, USA}

\author[label4]{Deying Xia}
\author[label4]{John A. Notte}

\address[label4]{Carl Zeiss SMT Inc, Danvers, MA 01923, USA}

\author[label1,label2]{Ruopeng Zhang}
\author[label1,label2]{Andrew M. Minor}

\author[label3]{Norman A. Sanford}
\ead{norman.sanford@nist.gov}

\begin{abstract}

We demonstrate a new focused ion beam sample preparation method for atom probe tomography.  The key aspect of the new method is that we use a neon ion beam for the final tip-shaping after conventional annulus milling using gallium ions. This dual-ion approach combines the benefits of the faster milling capability of the higher current gallium ion beam with the chemically inert and higher precision milling capability of the noble gas neon ion beam. Using a titanium-aluminum alloy and a layered aluminum/aluminum-oxide tunnel junction sample as test cases, we show that atom probe tips prepared using the combined gallium and neon ion approach are free from the gallium contamination that typically frustrates composition analysis of these materials due to implantation, diffusion, and embrittlement effects. We propose that by using a focused ion beam from a noble gas species, such as the neon ions demonstrated here, atom probe tomography can be more reliably performed on a larger range of materials than is currently possible using conventional techniques.

\end{abstract}

\begin{keyword}atom probe tomography \sep%
    focused ion beam \sep%
    gas field ionization source \sep%
    neon ions
\end{keyword}

\end{frontmatter}

\section{Introduction}

Quantification of the positions and identities of all constituent atoms in a material is a grand challenge of measurement science concerning a wide range of length scales of engineering interest. For example, in modern, nanoscale FinFET devices, a deficit or excess of only a few tens of dopant atoms can make or break performance \citep{Khan2016,Martin2018}. At the other extreme, catastrophic failure of jet engine turbine blades can arise from the nanoscale segregation of just a few atoms \citep{Rodenkirchen2022}. Therefore, precisely knowing the position and identity of constituent atoms down to the isotopic level is of paramount importance.

Atom probe tomography (APT) is at the forefront of the field of atomic quantification. Currently, APT yields sub-nanometer spatial resolution and a chemical sensitivity approaching a part per million \citep{Gault2021}. APT is well established for revealing intricate details about the chemical compositions of materials, and sophisticated instrumentation for APT continues to be developed \citep{Zhao2017,Chiaramonti2019,Vella2021,Prosa2021}. However, the ultimate success of the measurement hinges on the precise fabrication of the needle-like specimen, which must fulfill stringent requirements. Specimen fabrication is in fact still the major limiting factor for APT of many materials. 

The operation conditions for the APT measurement are harsh. In order to bring the electric field at the needle apex close to the threshold for field evaporation of ions (\unit[$\sim$5--50]{V/nm}), a large electrostatic stress is induced in the specimen due to the voltage bias applied. Ion emission is then triggered by thermal transients induced using a pulsed laser. The specimen must be sufficiently robust to avoid fracture during data acquisition and the key to success is to minimize the integrated stress, yet still apply an electric field strong enough to facilitate field evaporation.  However, the issue of electrostatic stress in the vicinity of the specimen apex can be quite complex, spatially discontinuous, and depend upon the dielectric, metallic, and semiconducting properties of the constituent layers that may be present \citep{Zhang2021}.  In practice, tip diameters of \unit[$<$100]{nm} are typically necessary in order to produce a sufficiently high electric field locally at the tip apex. Furthermore, the APT specimen should not be physically or chemically altered by the processing that forms the tips from the host material of interest, nor by the high electric field necessary for field evaporation. 

Although electropolishing schemes are routinely used to fabricate metal APT specimen tips \citep{Gault2012}, the most versatile method is focused ion beam (FIB) milling. In this process, a finely focused beam of ions, conventionally a gallium FIB, ``mills" the sample into a needle shape. Much has been accomplished using gallium FIB milling for preparing APT specimens \citep{Prosa2017}. However, there are some major pitfalls resulting from the use of gallium. Near the milled surface, gallium ions become implanted and in a crystalline sample can form an amorphous surface layer. This gallium contamination can propagate and accumulate further inside the specimen by diffusion along grain boundaries and interfaces, and/or by channeling between crystal lattice planes, inducing e.g.\ phase transformations in steels \citep{Babu2016}) and causing embrittlement in aluminum and its alloys \citep{Unocic2010}). Gallium can also alter the composition of e.g.\ group III-V compound semiconductors by preferential sputtering and surface diffusion effects \citep{Grossklaus2011}. 
 
Thus, the problem of gallium contamination and damage can severely frustrate APT analysis. While mass spectral data originating from the gallium-contaminated outer shell of the APT needle can sometimes simply be discarded from the subsequent 3D analysis, the effects of deeper infiltrations cannot, and they can even result in tip fracture during the APT run due to liquid-metal embrittlement. Often all of the APT data is required in a full 3D analysis and in that case discarding any subset is unacceptable. It has been shown that gallium penetration can be limited by performing the milling at cryogenic temperatures \citep{Lilensten2020,Macauley2021}. An alternative is to mill with a different ion species, preferably one that is chemically inert.

Over the last 15 years, several new FIB sources have been introduced (gas field-ionization, plasma, cold-atom, liquid-metal alloy, ionic liquid) offering a wider choice of ion species and beam parameters \citep{Smith2014,Gierak2018}. The ion microscope producing the smallest spot size for applications requiring the highest resolution is the helium ion microscope, which uses a specialized gas field ionization source (GFIS) \citep{Ward2006}. The source can also be operated with other gases, such as neon \citep{Livengood2011}. A recent review of nanofabrication applications using the helium(neon) ion microscope can be found in \cite{Allen2021}. The main reason to use the neon ion beam instead of the helium ion beam is the higher sputter yield of the heavier ion.

In the present work, we use neon FIB milling to avoid the gallium-induced limitations described previously, while at the same time improving the fabrication precision of the APT specimen. The maximum GFIS beam currents on the specimen are lower than those obtained using a gallium liquid-metal ion source (picoamps versus nanoamps), hence we still use conventional gallium FIB milling for the first fabrication steps and then switch to neon FIB milling using the GFIS for the final precision tip-shaping phase. The final neon milling step effectively removes any gallium-implanted layer. This approach builds on previous work in which a similar sequential approach was used to produce gallium-free lamella samples for transmission electron microscopy (TEM) by switching from gallium to neon FIB milling at a judicious timepoint \citep{Pekin2016}. 

As test specimens for the present study we focus on a titanium-aluminum alloy and a layered aluminum/aluminum-oxide tunnel junction sample, both of which can pose challenges for APT when using conventional gallium-only milling. Needle specimens are prepared using our sequential gallium and neon FIB milling approach, and control samples are prepared using gallium FIB milling only. The APT mass spectra and atomic reconstructions obtained from these samples are discussed and compared. We also discuss our neon FIB approach in the context of related work that has used xenon plasma FIB for APT sample preparation \citep{Estivill2016,Gault2018,Halpin2019,Famelton2021,Rielli2021}.

\section{Materials and Methods}
\label{Materials and Methods}

\subsection{FIB milling of needle-shaped specimens}

Neon and gallium FIB milling to prepare the needle-shaped specimens was performed using two Zeiss ORION NanoFab multibeam ion microscopes$^a$ interchangeably, the first at UC Berkeley and the second at Carl Zeiss SMT Inc. The ORION NanoFab instrument combines two ion columns: a conventional liquid-metal ion source (LMIS) column for gallium ions, and a GFIS column for helium or neon ions produced from ionization events at the apex of an atomically sharp tungsten tip. The nominal probe diameters of the He, Ne and Ga ion beams as measured in resolution tests using the rise distance method by the manufacturer are \unit[0.5, 2, and 3]{nm}, respectively. 

The substrates used were silicon coupons carrying arrays of flat-topped microtips onto which wedge-shaped sections of the material of interest (Ti-Al alloy and layered Al/AlOx structure) were platinum-welded following standard gallium FIB lift-out procedure for APT specimens \citep{Thompson2007}. A protective platinum layer was added to all samples prior to FIB milling; typically \unit[50]{nm} Pt by electron beam-induced deposition using an organometallic platinum-containing precursor followed by \unit[500]{nm} Pt by gallium ion-beam-induced deposition from the same precursor. The lift-outs were performed using an FEI Helios G4 FIB-SEM and a Zeiss Auriga FIB-SEM at LBNL and NIST-Boulder, respectively. 

\begin{figure*}[t!]
    \centering
    \includegraphics[width=0.9\linewidth]{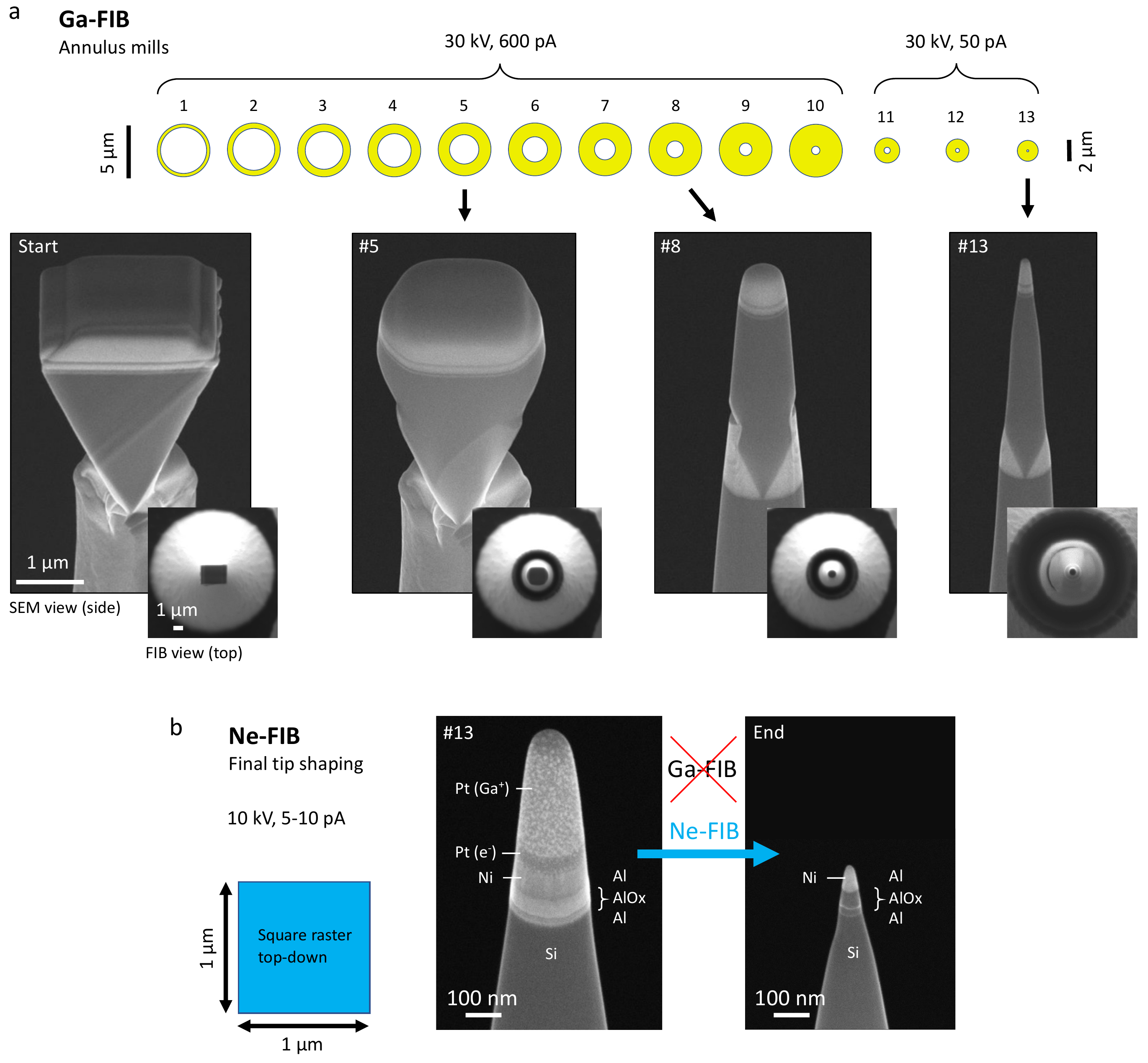}
    \caption{Sequential gallium and neon FIB milling strategy for fabricating needle-shaped APT specimens: (a) \unit[30]{kV} Ga-FIB milling sequence for annuli 1--10 (\unit[600]{pA}) starting with outer diameter \unit[5]{$\mu$m} and gradually decreasing the inner diameter from \unit[4.4--0.8]{$\mu$m}. For annuli 11--13 (\unit[50]{pA}) the outer and inner diameters decrease from \unit[2.4--2]{$\mu$m} and \unit[0.6--0.2]{$\mu$m}, respectively. (The complete set of milling parameters is given in the Supplementary Material.) (b) \unit[10]{kV} Ne-FIB final tip-shaping (\unit[5-10]{pA}) using a top-down square raster pattern centered on the tip apex and observing the contrast change in the secondary-electron neon ion microscopy (NIM) image in real time. The sample shown in the figure is the layered Al/AlOx sample and all images are scanning electron microscope (SEM) views acquired with a stage tilt of \unit[54]{$^\circ$} (SEM side views) and \unit[0]{$^\circ$} (equivalent to FIB top views). A protective platinum capping layer (\unit[$\ge$500]{nm} thick) covers the top surface of the wedge prior to milling.}
    \label{Milling_strategy}
\end{figure*}

In order to shape the wedges into the desired needle geometry, a series of top-down annulus mills with sequentially decreasing outer and inner diameters were implemented, as outlined in Figure \ref{Milling_strategy}a. These steps were performed with the gallium FIB of the ORION NanoFab, or using one of the FIB-SEM instruments. For the first ten annulus mills the beam acceleration voltage and current were set to \unit[30]{kV} and \unit[600]{pA}, respectively. For the last three annulus mills, the beam voltage was kept the same and the current was reduced to \unit[50]{pA}. The total mill time for the annulus patterns was $\sim$12 minutes.

\begin{figure}[t!]
    \centering
    \includegraphics[width=\linewidth]{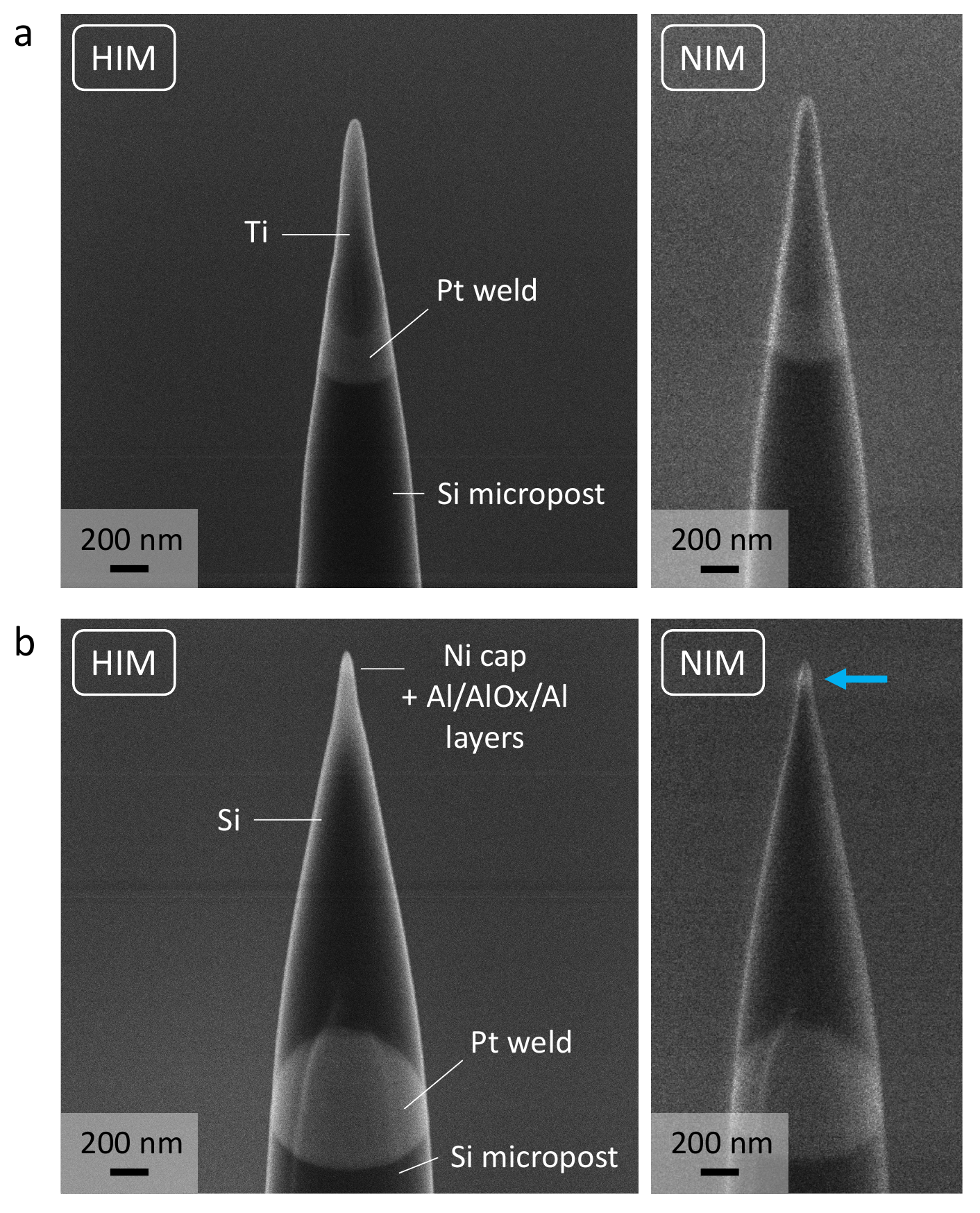}
    \caption{HIM and NIM images of APT needle specimens fabricated using the combined gallium and neon FIB method: (a) Ti-Al alloy sample. (b) Layered Al/AlOx sample; the arrow points to contrast from the Al/AlOx/Al stack. The stage tilt for these images was \unit[54]{$^\circ$}.}
    \label{HIM_NIM_images}
\end{figure}

The final tip-shaping step was performed as shown in Figure \ref{Milling_strategy}b, selecting a square shape of width \unit[$\sim$1]{$\mu$m} centered top-down on the needle and rastering the beam over that area (scan size 512$\times$512 pixels) while monitoring the secondary-electron image for several minutes (basically, this was just a ``reduced view" continuous scan). The size of the square was large enough not to interfere with the round shape of the tip. This milling step was performed using the neon FIB of the ORION NanoFab with a beam voltage of \unit[10]{kV} and a current of \unit[5--10]{pA}. For comparison, specimens were also prepared using gallium FIB milling for the final tip-shaping, using a \unit[5]{kV} gallium beam at a current of \unit[20]{pA}. It is typical to reduce the gallium beam acceleration voltage for this step in order to minimize the implantation depth of the gallium ions and the associated amorphization. By reducing the beam current, the milling rate is slowed hence allowing greater control. At the end of the tip-shaping, the protective platinum layer was completely removed and a needle with apex diameter generally in the range of  \unit[15--25]{nm} was obtained. Typical shank angles may be readily inferred from Figs.~\ref{Milling_strategy}  and \ref{HIM_NIM_images}.

Figure \ref{HIM_NIM_images} presents representative side-view helium ion microscopy (HIM) and neon ion microscopy (NIM) images of the two specimen types investigated in this work, both fabricated using the combined gallium and neon FIB method. These images were acquired directly after the final neon FIB polishing step shown in Fig.~\ref{Milling_strategy}b using a dwell time of \unit[1]{$\mu$s}, line average of 8, pixel spacing of \unit[1.5]{nm} (He) and \unit[4.8]{nm} (Ne), and a beam current of \unit[0.3--0.7]{pA}. 

In the case of the Ti-Al alloy sample (Fig.~\ref{HIM_NIM_images}(a)), the platinum capping layer that was added for protection during the gallium FIB lift-out procedure was (intentionally) entirely removed from the tip of the needle. For the Al/AlOx thin-film sample (Fig.~\ref{HIM_NIM_images}(b)), the platinum layer was again entirely removed, but a small portion of the nickel capping layer was left in place to serve as a parameter-tuning layer at the beginning of the APT run, since the critical layers of interest in this sample lie directly beneath. In the HIM/NIM images of this sample, the narrow band of bright contrast near the tip apex corresponds to the Al/AlOx/Al stack, residing just beneath the remaining nickel cap. It is interesting to note that the contrast from the Al/AlOx/Al stack is more evident in the NIM image than in the HIM image (as marked by the arrow in the NIM image of Fig.~\ref{HIM_NIM_images}(b)). While NIM resolution is inherently poorer than that of HIM (due to the larger effective probe size of the neon beam), the microscope operator can still use NIM imaging to assess neon FIB milling progress without needing to switch to the helium ion beam. And as shown here, contrast from the layers of interest may even be enhanced in the NIM mode.

\subsection{APT measurements}

APT was performed using the CAMECA LEAP 4000X Si APT instrument at NIST-Boulder. Readers unfamiliar with the basic operational principles and common terminology of APT are referred to \cite{Larson2013}. For all APT work described herein, the needle specimens were held at a base temperature of \unit[$\sim$50]{K} in ultrahigh vacuum during analysis. The \unit[355]{nm} laser installed on the instrument was pulsed at a repetition rate of \unit[250]{kHz}, and the flight length between specimen and detector was \unit[90]{mm}. Other more specimen-specific details are now given.

\subsubsection{Titanium-aluminum alloy}

Successful APT runs were performed on six tips out of a batch of eight that were entirely gallium FIB milled and polished. Of these eight tips, one fractured during APT and the remaining tip would not exhibit field evaporation. Thus, the specimen yield for this small batch of eight gallium-milled tips was \unit[75]{\%.} Ten tips were prepared by combined gallium followed by neon FIB, resulting in nine successful APT runs; the tenth tip displayed a highly asymmetric morphology and was not attempted.  Therefore, the nine successful attempts on nine properly milled tips represent a \unit[100]{\%} yield for the combined Ga-Ne FIB processing. We note that while these were the success rates for the sample sets examined, they represent fairly low populations from which to infer any strong statistical conclusions.

For comparing APT outcomes from the two FIB preparation methods, we present results where a tip from each set ran without fracture to a standing voltage in the range of \unit[2.5--5.8]{kV}, roughly 10 million ions were recorded, the incident laser pulse energy was \unit[30]{pJ}, and the portion of each tip lost to field evaporation could be estimated by imagery before and after APT.  Finally, it is informative to mention the approximate magnitude of the electric field strength at the apex of our example APT tips that were run at \unit[30]{pJ}. Considering the tip apex diameters before and after APT analysis and the span of the standing voltages over the course of the data acquisition runs, we can draw from the results of \cite{Zhang2021} and \cite{Larson2013} to estimate that the (apex) surface electric field on these specimen tips was in the approximate range of \unit[20--50]{V/nm}.

\subsubsection{Aluminum/aluminum-oxide multilayer}

For this sample type, nine tips were prepared entirely by gallium FIB milling and all of those were prone to fracture and ran very poorly in APT. Surviving tips yielded mass spectra that showed significant gallium contamination and no useful distinction of the constituent layers. Hence the specimen yield for the gallium-only tips was essentially \unit[0]{\%}. On the other hand, of the nine tips prepared by gallium FIB milling followed by neon FIB polishing, two ran successfully in terms of showing mass spectra representative of the nickel, aluminum, and titanium layers, and revealed no gallium contamination. Thus, if the success metric is regarded as only detecting nickel, aluminum, and titanium, and finding no gallium, then the specimen yield is \unit[22]{\%}. Notably, one of these ``successful" tips ran through the entire layer to the silicon substrate without catastrophic fracture and we abbreviate this tip as TJ-1. The other tip (TJ-2) ran through the layer stack but fractured at the Ti/Si interface and was destroyed. However, both tips showed jumps in the detection rate at the Ni/Al interface suggesting at least partial fracture in the Ni/Al interface region followed by the tips re-forming and continuing to run through the aluminum and into the titanium layer. The tip that ran through the layer stack and into the silicon substrate without suffering catastrophic fracture offers plausible evidence for APT detection of the Al/AlOx tunnel junction, as described in more detail below.

\subsection{Complimentary TEM analysis}

TEM characterization of the samples investigated in this work was performed using an FEI TitanX microscope (LBNL).

In the case of the Ti-Al alloy sample, TEM analysis focused on examining any milling artifacts from the neon FIB. For this, a lamella was extracted from the bulk sample by conventional gallium FIB lift-out (lamella thickness \unit[2]{$\mu$m}) and platinum-welded onto a TEM half-grid. A series of needle-shaped specimens was formed in the top side of the lamella using gallium FIB annulus milling and neon FIB polish, following the milling procedure shown in Fig \ref{Milling_strategy}. The TEM images of these specimens were acquired at an acceleration voltage of \unit[300]{kV}. 

For the layered Al/AlOx material, scanning TEM (STEM) inspection was used to check the relative thicknesses of the layers. A TEM lamella was prepared by gallium FIB lift-out and thinned to \unit[$\sim$100]{nm} by gallium FIB milling using a \unit[5]{kV} beam with current \unit[30]{pA} in the final milling step. Elemental mapping was performed by STEM-based X-ray energy-dispersive spectrometry (XEDS) at \unit[300]{kV} using a Bruker Super-X quadrature detector (solid angle \unit[0.7]{sr}). The STEM-XEDS data were analyzed using Bruker Esprit software.

\section{Results and Discussion}

\subsection{Titanium-aluminum alloy}

\begin{figure*}[t!]
    \centering
    \includegraphics[width=\linewidth]{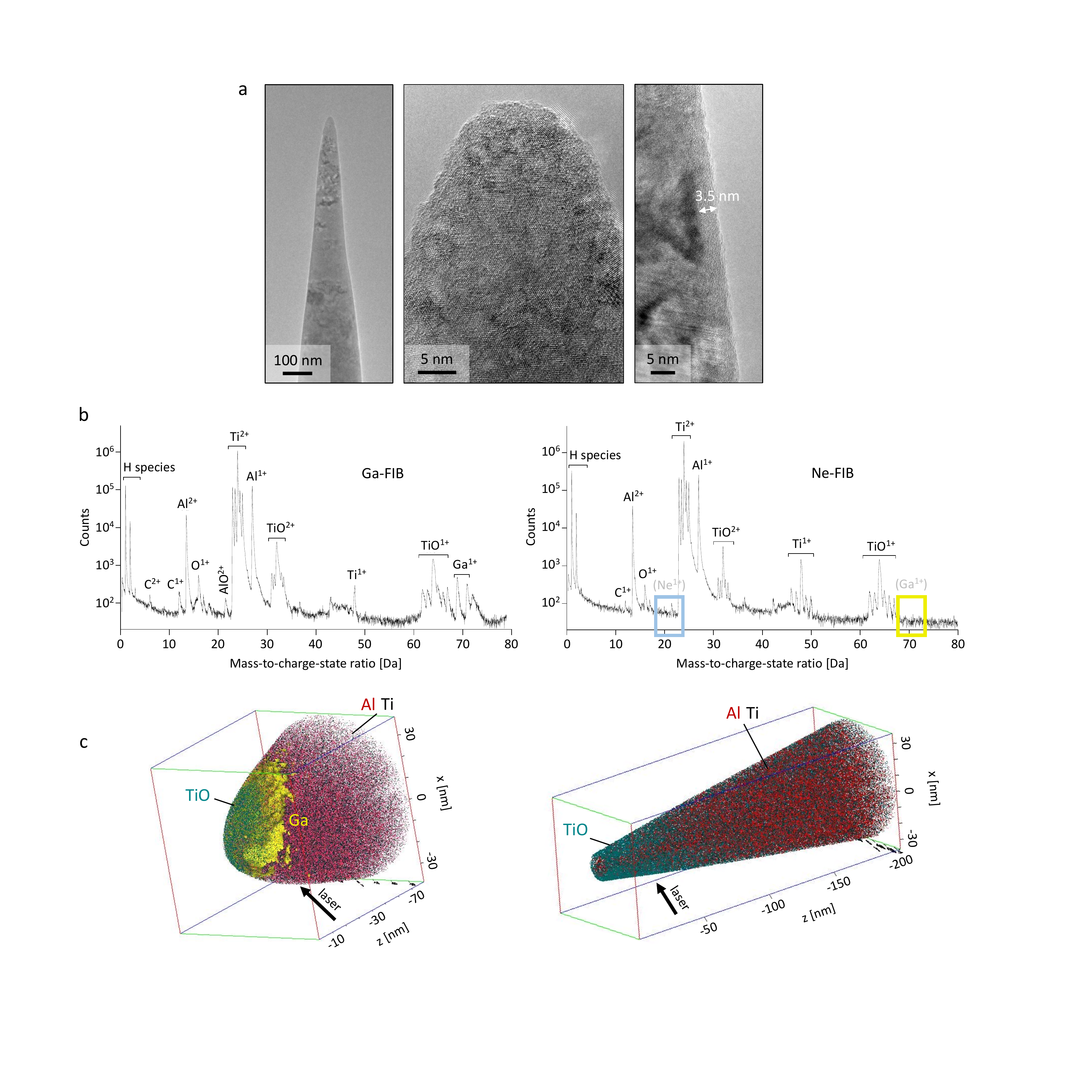}
    \caption{Results for the Ti/Al alloy sample: (a) TEM images of needle specimen produced using the gallium followed by neon FIB milling approach, showing a high aspect ratio (left), a tip apex diameter of \unit[$\sim$20]{nm} (middle), and a \unit[$\sim$3.5]{nm} surface layer (right). (b) APT mass spectra obtained from a specimen milled using gallium FIB only (left) compared with a specimen milled using neon FIB for the final polish (right). The spectral regions where peaks from implanted gallium and neon ions would be expected if present (assuming singly charged species) have been highlighted. (c) APT 3D reconstructions for the respective mass spectra shown above: Al (red), Ti (black), TiO (dark green), Ga (mustard yellow). Ga is shown as an isoconcentration surface at \unit[0.5]{at.$\%$} while all other constituents appear as data points. The Al and Ti counts have been weighted to highlight the distribution of the minor component in the alloy (Al).}
    \label{Ti-Al_results}
\end{figure*}

The results for the Ti-Al alloy sample are presented in Fig.~\ref{Ti-Al_results}. The particular alloy investigated here is of interest due to localized deformation behavior observed upon annealing, which is hypothesized to arise due to short-range ordering \citep{Zhang2019}. 

First we consider the TEM analysis that is shown in Fig.~\ref{Ti-Al_results}(a). Here we see that using the \unit[10]{kV} neon FIB polishing strategy, a sharp needle with crystallinity preserved up to the outer few nanometers is achieved. A surface layer of thickness \unit[$\sim$3.5]{nm} (with both amorphous and crystalline character, see Supplementary Fig.~S1) is measured. The surface layer thickness corresponds well with the simulated penetration depth for \unit[10]{kV} neon ions and the associated vacancy depth distribution, as discussed further in Section 3.3. The TEM analysis also demonstrates the precise milling capability of the neon FIB, with resulting apex diameters comparable to what is obtainable with conventional gallium FIB. For the tip shown in Fig.~\ref{Ti-Al_results}(a), an apex diameter of \unit[20]{nm} is measured. Compared with the recently introduced xenon plasma FIB method \citep{Estivill2016,Gault2018}, we believe that the neon FIB method enables finer control over the final tip shape due to the lower beam current and lower sputter yield of the lighter ions. Conversely, an advantage of the xenon plasma FIB is that it can be used to mill larger volumes of material to produce APT tips from bulk samples without the need for a lift-out \citep{Halpin2019,Famelton2021,Rielli2021}. 

In Fig.~\ref{Ti-Al_results}(b), the APT mass spectra obtained from a control specimen produced using gallium FIB milling only (left) and from a specimen polished using the neon FIB in the final step (right) are compared. In both cases, the respective, full data sets were used and have not been parsed to isolate ions emitted from spatially distinct subsets within the sample volumes examined. It becomes clear that the neon FIB approach has produced a specimen with no detectable gallium residue, indicating that by switching ion species from gallium to neon near the end of the milling sequence, a gallium-free specimen can be produced. Furthermore, for the neon FIB specimen, no residual neon was detected either. This could be due to out-diffusion of the noble gas species over the shallow implantation depth, or any residual neon could be below the APT detection limit. There are some distinguishable differences in the titanium and oxygen peaks between the two specimens. These distinctions presumably arise from the different FIB preparation methods, however, as discussed further below, they have negligible impact on APT estimates of the alloy fraction in each case.

When comparing APT quantification of the two specimens, the field of view (FOV) of the detector must be considered. For example, for the case of the all-gallium FIB processing, a gallium implantation layer will be present on the entire surface of the FIB-processed tip, but only the subset of field-evaporated ions that fall within the detector's FOV are collected and available for reconstruction analysis. Therefore, as ions are removed and the apex diameter of the conical tip increases, the gallium-implanted surface of the specimen will depart from the FOV. An example of this is illustrated in Fig.~\ref{Ti-Al_results}(c) (left), which shows a gallium concentration near the nose of the reconstruction and vanishing gallium elsewhere. Therefore, in comparing APT-derived estimates of composition for the two cases, the analysis volumes are constrained to regions of interest (ROIs) that deliberately exclude the surface and apex region of the physical tip to thus avoid counting surface gallium and other surface-related contaminants. 

In addition to constraining the ROI analysis volumes as just described, quantification of the mass spectra must be accompanied with caveats associated with the background subtraction scheme used to scale the counts. The supplied specifications of the sample give an aluminum alloy fraction of \unit[9.8]{at.$\%$}. Both the global-time-of-flight (GTOF) and local-range-assisted (LRA) background subtraction methods bundled into the CAMECA IVAS analysis software return aluminum concentrations of \unit[9.9]{at.$\%$} for the neon FIB polished specimen. However, for the all-gallium milled specimen the respective GTOF- and LRA-derived aluminum concentrations are \unit[9.8]{at.$\%$} and \unit[9.5]{at.$\%$}, respectively.  

Hence, in terms of compositional quantification, both milling strategies yielded comparable results, but foreknowledge of the sample composition allows one to conclude that the GTOF background subtraction method is preferred in this case. Besides simply the presence or absence of gallium, the FIB processing techniques may also differ in the robustness of the needle specimens. In the case of materials that are susceptible to gallium embrittlement~\citep{Unocic2010}, premature fracture during APT data acquisition often occurs when the conventional gallium FIB method has been used for specimen preparation. That trend may in fact exist in the present case but we do not have enough trials to make a strong statistical argument.

Additional details involving the 3D reconstructions deserve discussion. Referring again to Fig.~\ref{Ti-Al_results}(c) (left), the specimen tip did not suffer fracture during APT so the length of material removed was measured and used as input to scale the reconstruction parameters. Beyond that, however, use of the imported shape of the tip (prior to APT) to compute the reconstruction via the ``Tip Profile" method did not yield geometrically satisfactory results. The ``Shank" scheme was not applicable since the tip was of rather blunt morphology and did not display a uniform shank angle.   Therefore, the reconstruction displayed in Fig.~\ref{Ti-Al_results}(c) (left) was computed by means of the ``Voltage" method. On the other hand, the Tip Profile method did reconstruct a realistic tip shape for the combined gallium  and neon FIB processing, as shown on the right side of Fig.~\ref{Ti-Al_results}(c). For that case, however, the TiO species are not seen to encase the entire nose of the tip because counts comprising the actual apex were consumed in aligning the instrument and were not included in the reconstruction.

APT of the gallium-FIB-only specimen revealed gallium implantation residue appearing as a rotationally (with respect to the z axis) asymmetrical layer near the reconstruction apex (Fig.~\ref{Ti-Al_results}(c) (left)). The asymmetry is correlated with the incident laser direction where the shadow side of the tip reveals a higher level of apparent gallium residue (and TiO) than the illuminated side. Of course, from the standpoint of the FIB processing, such an implantation asymmetry is quite improbable and the observed effect is most likely an artifact of APT -- possibly resulting from nonuniform tip heating associated with the laser illuminating one side of the tip. Indeed, artifacts associated with laser input direction are not unexpected in APT~\citep{Shariq2009}, and tip illumination schemes involving multiple laser input directions have been invented~\citep{Bunton2022}.

\subsection{Aluminum/aluminum-oxide multilayer}

\begin{figure*}[p]
    \centering
    \includegraphics[width=\linewidth]{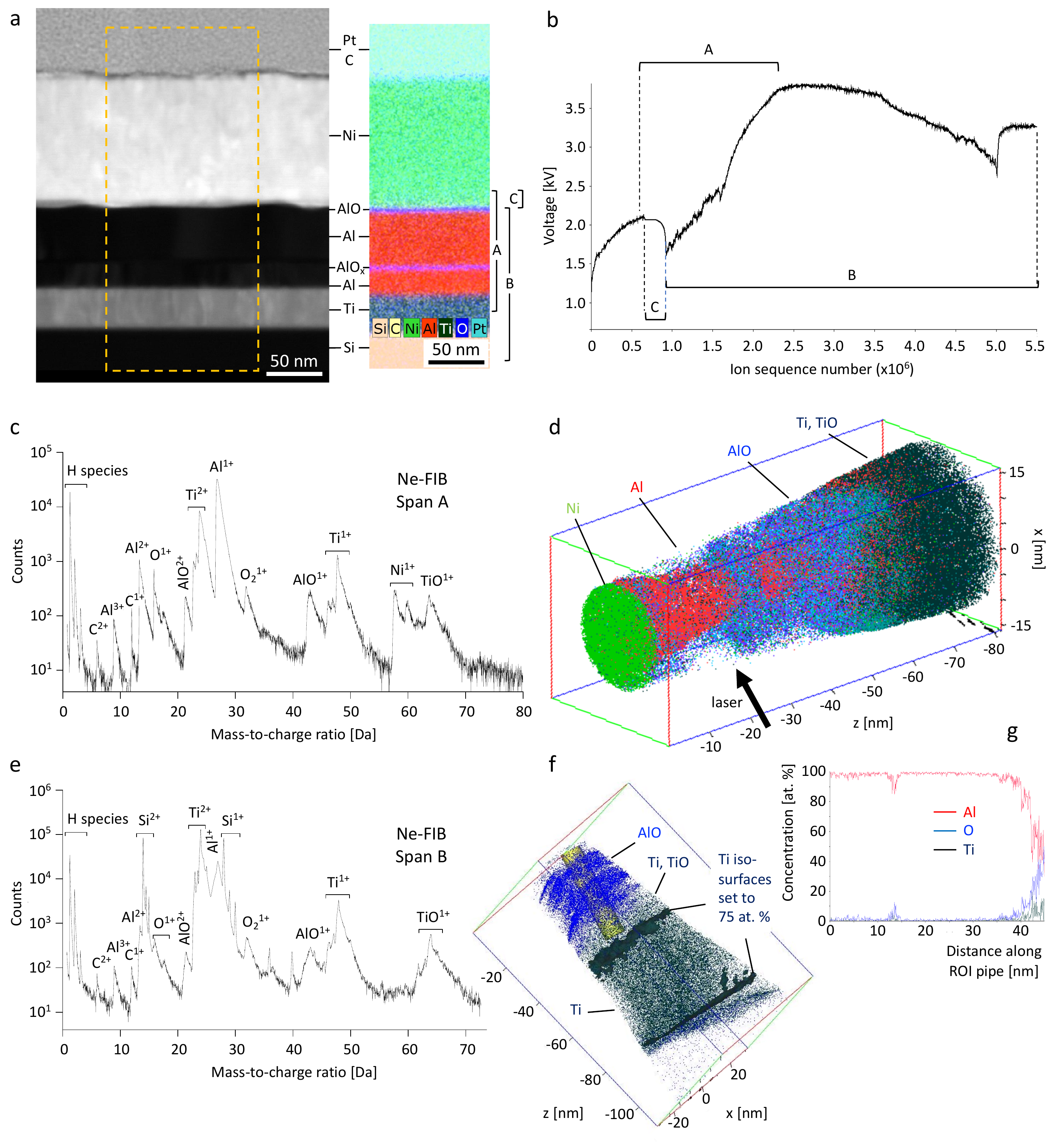}
    \caption{Results for the Al/AlOx multilayer sample: (a) STEM image (left) and STEM-XEDS elemental map (right) showing the sample layers including the silicon substrate and the protective platinum/carbon layer added before the FIB milling of the cross section. The portions marked with the square brackets on the far right show the approximate locations of the regimes interrogated by APT as discussed subsequently. (b) Voltage history and ion sequence number from the APT run of sample TJ-1 milled using neon FIB for the final polish with three distinct regimes marked: span A (encompassing a portion of the nickel layer, all of the Al/AlOx/Al stack, and a portion of the titanium layer), span B (encompassing the Al/AlOx/Al stack, titanium layer, and a portion of the silicon substrate), and span C (portion of the nickel layer where a transient jump in detection rate suggests partial fracture at the Ni/Al interface). (c) APT mass spectrum corresponding to span A. (d) 3D reconstruction for span A. (e) APT mass spectrum corresponding to span B. (f) 3D reconstruction for span B; an ROI pipe of diameter \unit[10]{nm} has been placed in the aluminum layer, tilted to account for the specimen tilt, and offset to avoid the oxidized surface of the aluminum. To improve image clarity, aluminum and silicon counts have been omitted. (g) Concentrations of aluminum, oxygen and titanium along the ROI pipe; the feature at a depth of \unit[$\sim$12]{nm} suggests that the AlOx tunnel junction has been resolved. We note that gallium-FIB-only specimens yielded no useful data due to premature fracture during the APT experiment.}
    \label{Al-AlO_results}
\end{figure*}

The results for the second test sample are presented in Fig.~\ref{Al-AlO_results}. First we show the baseline STEM and STEM-XEDS characterization of the thickness and composition of the layers (Fig.~\ref{Al-AlO_results}a). At the base is the silicon substrate, with a titanium layer of thickness \unit[$\sim$36]{nm} on top. The titanium layer also contains oxygen and a degree of carbon. Next comes the Al/AlOx/Al stack composed of \unit[$\sim$22]{nm} and \unit[45]{nm} aluminum layers sandwiching the AlOx tunnel junction layer of thickness \unit[3--4]{nm}. The upper aluminum layer also has an oxygen-rich layer on top (\unit[$\sim$5]{nm}). Next is the nickel cap of thickness \unit[$\sim$120]{nm}, added to serve as a sacrificial parameter-tuning layer at the start of the APT run. Finally there is a platinum/carbon layer, added during the FIB lift-out procedure to protect the underlying material from damage from the FIB.

The ultimate metrology challenge in the analysis of such tunnel junction structures is to measure the oxygen concentration in the oxide layer, since the nature of the oxide is especially critical to device performance \citep{Zeng2015}. APT is well suited for this task since it offers sub-nanometer spatial resolution, and with caveats (including a suitably large analysis volume and sufficiently low background within the m/Z mass spectral range of analytical interest) a chemical sensitivity in the tens of parts-per-million range. Practically speaking, however, given the junction thickness as revealed by STEM, the specimen tip dimensions, and the available FOV of the APT detector, we will certainly not achieve such low sensitivity for quantification of the oxygen concentration in the tunnel junction in our present experiments. Nonetheless, in the discussion that follows, we show that APT plausibly reveals the tunnel junction layer for the more successful of the two specimen tips (TJ-1) that was prepared by combined gallium and neon FIB.  The entirely gallium-FIB-milled APT specimens of the tunnel junction suffered severe corruption due to gallium implantation, fractured easily, yielded no useful data, and are not considered further.

A detailed discussion of the APT analysis for specimen TJ-1 is now presented: A relatively high incident laser pulse energy of \unit[50]{pJ} was used. This triggered field evaporation for a tip-electrode voltage varying from \unit[1.5--3.8]{kV} over the course of the run. Such a relatively low tip-electrode voltage helped reduce the probability of tip fracture, but the elevated laser pulse energy promotes sample heating and also ``tails" in the mass spectra, which tend to increase background and reduce the fidelity and distinction between closely spaced peaks. The voltage history versus ion sequence number is shown in Fig.~\ref{Al-AlO_results}(b) and three distinct regimes spanning different portions of ion sequence number are identified. The approximate spatial locations of these three regimes in the sample stack are marked in Fig.~\ref{Al-AlO_results}(a). Span A encompasses a portion of the nickel cap, both aluminum layers, and a portion of the titanium layer; the mass spectrum and reconstruction corresponding to span A are shown in Figs.~\ref{Al-AlO_results}(c) and (d), respectively. Span B omits the nickel layer, encompasses the aluminum and titanium layers, and also includes a portion of the silicon substrate; the respective mass spectrum and reconstruction for span B are illustrated in Figs.~\ref{Al-AlO_results}(e) and (f). Notably, as the Ni/Al interface is penetrated, the detection rate over span C jumps abruptly from the target value of \unit[0.3]{\%} up to roughly \unit[100]{\%}, before settling back to \unit[0.3]{\%} as APT into the aluminum layer progresses.

The reconstruction shown in Fig.~\ref{Al-AlO_results}(d) was computed under the assumption that the Ni/Al and Al/Ti interfaces should be essentially flat and parallel. However, no amount of adjusting the reconstruction parameters while holding the (predetermined) layer thicknesses fixed would yield such a result. The strong jump in detection rate noted in span C of Fig.~\ref{Al-AlO_results}(b) suggests that the Ni/Al interface has suffered at least a partial fracture, which would leave the Ni/Al interface indeterminate and help explain the inability of the IVAS software to simultaneously reconstruct the Ni/Al and Al/Ti interfaces.

With the foregoing points in mind, the reconstruction shown in Fig.~\ref{Al-AlO_results}(f) is computed under the assumption that the nickel layer and a portion of the aluminum layer have been fractured away, and is therefore scaled to the measured (by STEM) thickness of the titanium layer. The titanium layer is bracketed by isoconcentration (iso) surfaces and the reconstruction has been rotated and oriented to view the iso surfaces edge-on such that they roughly appear as lines. As shown in the figure, this scenario reveals that the lift-out was mounted under a tilt of roughly \unit[10]{$^\circ$}. Therefore, if the AlOx junction layer is to be resolved, an ROI must be tilted to compensate for the FIB mounting error. The reconstruction also reveals that the surface of the aluminum portion of the tip is highly oxidized. Therefore, the ROI should also exclude the oxidized surface of the aluminum region since otherwise the contrast of the junction AlOx layer with respect to the purely aluminum layers would be lost.

In order to probe for the presence of the AlOx junction layer, a \unit[10]{nm} diameter ROI pipe has thus been placed in the aluminum portion of the reconstruction shown in Fig.~\ref{Al-AlO_results}(f). Figure~\ref{Al-AlO_results}(g) shows the resulting axial concentration profile for aluminum, oxygen, and titanium along the length of the pipe. As described above, the pipe is positioned to largely exclude the oxidized surface of the aluminum layer and is tilted to best reveal a feature in the composition profile, which suggests resolution of the junction layer, occurring at a depth of \unit[11--14]{nm} from the nose of the reconstruction. Near its maximum depth, the tilted ROI pipe encounters the oxidized sidewall of the aluminum layer and begins to intrude on the titanium layer. The tilt of the ROI pipe required to best reveal the presumed tunnel junction roughly conforms to the reconstruction-determined tilt of the FIB mount.  Additionally, the separation of the presumed tunnel junction and onset of titanium counts essentially conform to the \unit[22]{nm} thickness of the bottom aluminum layer as measured by STEM. Therefore, within the scenario outlined, the partial fracture on the tip at the Ni/Al interface did not also remove the tunnel junction layer and it is quite plausible that IVAS has revealed the tunnel junction layer as described. From this APT analysis, the fraction of oxygen in the tunnel junction layer is determined to be in the range of \unit[10--15]{at.\%}. This is in good agreement with the oxygen fraction that was computed from the STEM-XEDS map, which was \unit[$\sim$11]{at.\%}. Moreover, and within uncertainty, the thickness of the junction layer revealed by APT is also in agreement with the STEM-XEDS.

\begin{figure*}[t!]
    \centering
    \includegraphics[width=0.8\linewidth]{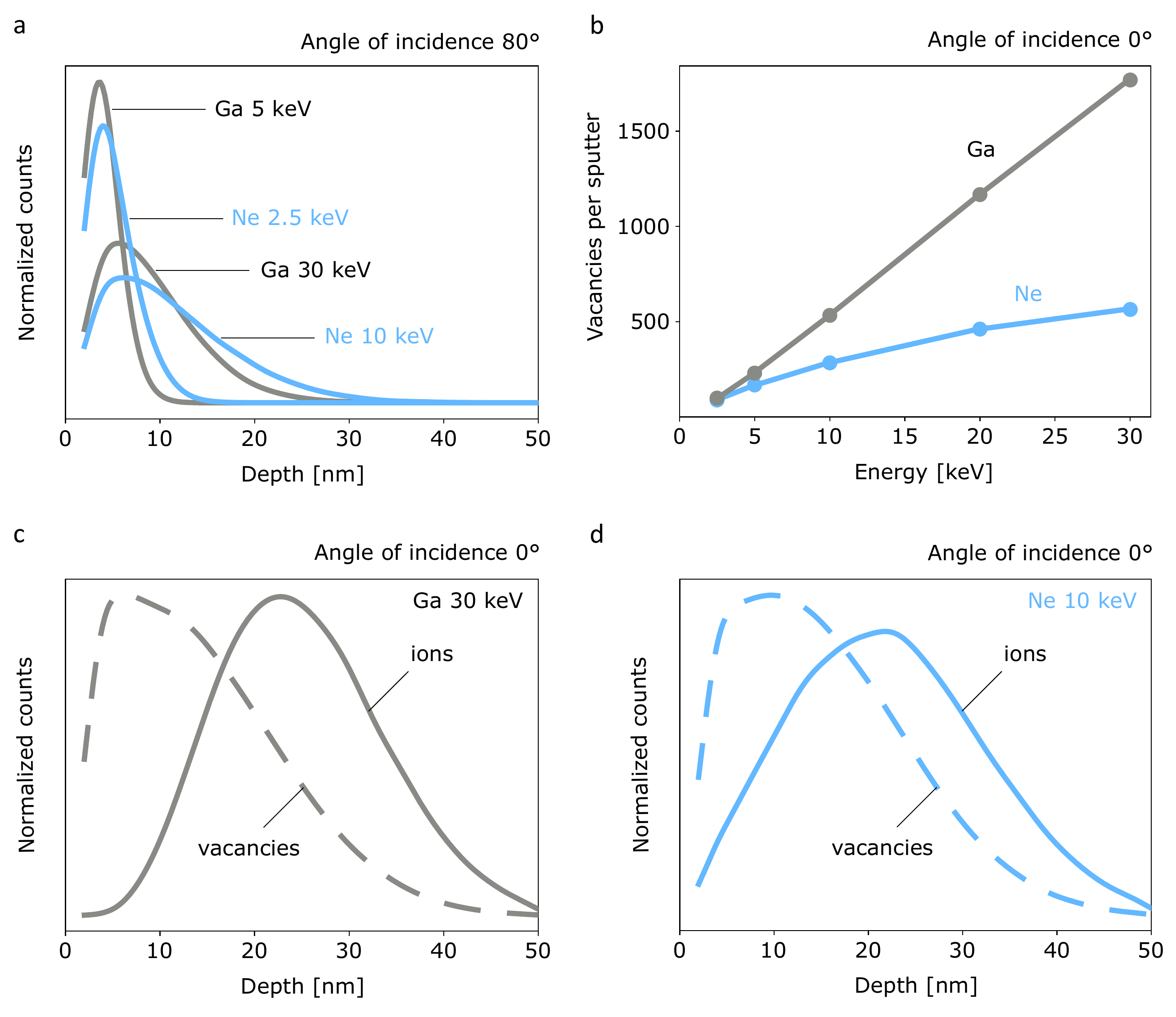}
    \caption{Monte Carlo simulations of ion stopping and vacancy generation for an aluminum target: (a) Gallium ion (\unit[5 and 30]{kV}) and neon ion (\unit[2.5 and 10]{kV}) depth distributions for an angle of incidence of \unit[80]{$^\circ$}, normalized by total intensity. (b) Number of vacancies generated per sputtered atom versus beam energy for gallium and neon ions, for an angle of incidence of \unit[0]{$^\circ$}. (c) and (d) Depth distributions for \unit[30]{kV} gallium ions and \unit[10]{kV} neon ions, respectively, compared with the corresponding vacancy depth distributions, for an angle of incidence of \unit[0]{$^\circ$}.}
    \label{Simulations}
\end{figure*}

The extremely low yield and fragility of even the gallium/neon milled tunnel junction specimens suggest further exploratory work to improve outcomes. For example, the excessive oxidation of the aluminum sidewalls of the tips could be prevented by an inert environment ``suitcase" transfer between FIB and APT tools. Additionally, oxide interfaces may be better characterized in APT by operating at laser wavelengths shorter than the \unit[355]{nm} used herein. Notably, in their APT study of Si/SiO$_2$(\unit[12]{nm})/Si test structures, \cite{Prosa2021} found improved results in rendering the oxide interface when using a laser operating at \unit[266]{nm} compared to \unit[355]{nm}.  Finally, we  also note that new APT instrumentation, which employs an even deeper UV source (\unit[29.6]{nm}), may offer additional advantages for examining oxides~\citep{Chiaramonti2020}.     

\subsection{Comparison of gallium and neon ion damage distributions from simulations}

To consider in more detail what determines the thickness of the surface layer observed by TEM for the APT needles (Fig.~\ref{Ti-Al_results}(a) far right, and Supplementary Fig.~S1), we have performed Monte Carlo simulations using the Stopping and Range of Ions in Matter (SRIM) code developed by \cite{Ziegler2010}. This allows us to compare the stopping range for gallium and neon ions and the effect of ion energy thereon (Fig.~\ref{Simulations}(a)), as well as the effect of ion species and energy on the number of vacancies generated per sputter (Fig.~\ref{Simulations}(b)). In addition, we compare the vacancy depth distributions with the corresponding ion stopping ranges for both ion species in  Figs.~\ref{Simulations}(c) and (d). The target material used for these simulations is aluminum. By default, ion channeling effects, which are dependent on crystal orientation, are neglected in the SRIM code.

The angle of incidence chosen for the ion stopping range simulations in Fig.~\ref{Simulations}(a) is \unit[80]{$^\circ$} with respect to the sample normal, reflecting the geometry of the vertical ion beam impinging on the slanted side of the needle-shaped specimen. The gallium ion beam potentials are selected to be \unit[30 and 5]{kV}, since these are typical parameters used for the annulus milling and final tip-shaping steps, respectively, when following conventional gallium FIB protocol. From the plot it can be seen that as the beam energy is reduced, the stopping distance of the ions becomes shorter. This is the reason why lower beam energies are used in the final polishing steps for both APT specimens and TEM lamellae, since a shorter stopping range means a thinner damage layer.

The neon beam potentials used for the simulation in Fig.~\ref{Simulations}(a) were \unit[10 and 2.5]{kV}, which give depth distributions that closely mirror those of the \unit[30 and 5]{kV} gallium ions. Indeed, as has been shown previously, \unit[10]{kV} neon ions can be used to approximate the depth distribution of \unit[30]{kV} gallium ions \citep{Pekin2016}. In order to decrease the thickness of the damage layer, a lower beam energy for the neon ion beam milling would be preferable, but was prevented in this work by instrument limitations. Nevertheless, with a simulated depth distribution peaking at \unit[5--6]{nm} for \unit[10]{kV} neon ions impinging on the aluminum target, amorphization is expected to be confined to the outer few nanometers of the APT needles. When simulating for a titanium target (the main component of the alloy) very similar results are obtained, as shown in Supplementary Fig.~S2.

In Fig.~\ref{Simulations}(b), we consider an angle of incidence of \unit[0]{$^\circ$} with respect to the sample normal, which approximates the beam-sample geometry at the tip apex. The plot indicates that the number of vacancies generated per sputter increases with beam energy, increasing more steeply for gallium ions compared to neon ions. Since the number of vacancies generated per sputter is directly related to amorphization, this suggests that for the same number of sputtered atoms (i.e.~for the same needle shape), there will be less subsurface damage from neon ions than from gallium ions when comparing the same beam energy. 

Moving on to Figs.~\ref{Simulations}(c) and (d), we consider the depth distributions of the ions together with the vacancies they generate for \unit[30]{kV} gallium and \unit[10]{kV} neon ions, respectively (angle of incidence \unit[0]{$^\circ$}). We observe that the stopping depth of the ions peaks deeper in the sample than that of the ion-induced vacancies. Furthermore, when considering the beam-sample geometry on the sides of the needle (angle of incidence \unit[80]{$^\circ$}), we find that most of the vacancies are generated within the first few nanometers (Supplementary Fig.~S3). This is in good agreement with the TEM analysis of the \unit[10]{kV} neon FIB polished tip, for which a partially amorphized surface layer of thickness \unit[3.5]{nm} was measured (Fig.~\ref{Ti-Al_results}(a), Supplementary Fig.~S1). Similar surface layer thicknesses have been experimentally observed for \unit[30]{kV} gallium FIB milled tips \citep{Zhong2020}. Polishing with the heavier xenon ions (shortest stopping distance) using a plasma FIB, implementing beam potentials down to \unit[2]{kV}, has been shown to produce amorphous layers with thicknesses under \unit[2]{nm} \citep{Halpin2019}. As has been noted previously, such surface layers can also involve native oxide and/or contaminants (oxygen and carbon) adsorbed during the milling process \citep{Estivill2016}.

\section{Conclusions}

In summary, we have shown how a neon FIB can be used in combination with conventional gallium FIB milling to produce APT specimens that are free from the gallium contamination that can severely hamper APT analysis of a range of materials. The initial annulus milling steps are performed with the gallium FIB and then the neon FIB is used for the final polish, thereby removing gallium residue from the previous annulus mill sequence and allowing precise shaping of the final tip. This dual-ion approach combines the benefits of the higher milling rates of the gallium FIB with the chemical inertness and greater precision of the neon FIB. Demonstrating for a titanium-aluminum alloy and for a multilayer aluminum/aluminum-oxide structure, we have shown that the new method produces pristine APT specimens that generally run to completion in the APT experiment without premature fracture due to gallium embrittlement, and which can be fabricated within a reasonable time frame. 

For APT of materials that are particularly affected by gallium contamination (e.g.\ aluminum and its alloys, group III-V semiconductors, and various thin-film structures), we thus propose the method introduced herein. Continuing development of noble gas ion sources together with advances in beam deceleration to minimize surface amorphization can further enhance the approach. Techniques that allow flexible preparation of APT specimens will also promote the development of APT instrumentation itself. For example, in the case of ion emission triggered using ultrafast extreme-UV (EUV) excitation, the EUV photons are strongly absorbed within the volume of the illuminated tip regardless of composition \citep{Chiaramonti2019,WinNT}. Hence the method used to prepare the specimen tip can have a strong influence on the EUV/tip interaction and the resulting field evaporation of ions. Comparative analyses of tips that have been prepared by both the gallium and neon FIB methods could thus be used to better understand and optimize the EUV-induced field evaporation process.      

\section*{Acknowledgements}

This work was funded in part by NIST. R.Z.~and A.M.M.~thank the U.S.~Office of Naval Research for funding under Grant No.~N00014-19-1-2376. Work at the Molecular Foundry at Lawrence Berkeley National Laboratory was supported by the Office of Science, Office of Basic Energy Sciences, of the U.S. Department of Energy under Contract No.~DE-AC02-05CH11231. We thank Soeren Eyhusen of Carl Zeiss for technical insights during the early stages of this project. We gratefully acknowledge the insight of a reviewer who suggested examining the scenario involving partial fracture of the Ni/Al interface in the Al/AlOx tunnel junction sample. 

\section*{Author notes}

At the time of this research, two of the authors (JAN, DX) were employed by Carl Zeiss, Inc., which manufactures the ORION NanoFab product used in this research

\section*{Footnotes}

$^a$Commercial equipment, instruments, or materials are identified only in order to adequately specify certain procedures. In no case does such identification imply recommendation or endorsement by the National Institute of Standards and Technology, nor does it imply that the products identified are necessarily the best available for the purpose.

\newpage

\bibliographystyle{elsarticle-harv}

\bibliography{bib.bib}

\end{document}